\documentclass[prl,twocolumn]{revtex4-1}
\usepackage{graphicx,bm,float,color,hyperref,amssymb,amsmath,physics}

\newcommand{\al}{\alpha}

\newcommand{\de}{\delta}
\newcommand{\D}{\Delta}
\newcommand{\e}{\epsilon}

\newcommand{\g}{\gamma}

\newcommand{\w}{\omega}

\newcommand{\s}{\sigma}

\newcommand{\mbJ}{\mathbf{J}}
\newcommand{\mbk}{\mathbf{k}}
\newcommand{\mbK}{\mathbf{K}}

\newcommand{\mbn}{\mathbf{n}}

\newcommand{\mbq}{\mathbf{q}}
\newcommand{\mbQ}{\mathbf{Q}}
\newcommand{\mbr}{\mathbf{r}}

\newcommand{\mbS}{\mathbf{S}}

\newcommand{\bsphi}[1]{\boldsymbol{\phi}}

\newcommand{\mcal}[1]{\mathcal{#1}}

\newcommand{\nn}{\nonumber\\}

\newcommand{\ben}{\begin{equation}}
\newcommand{\een}{\end{equation}}
\newcommand{\bens}{\begin{subequations}\begin{align}}
\newcommand{\eens}{\end{align}\end{subequations}}
\newcommand{\eq}[1]{Eq.~\eqref{#1}}
\newcommand{\fig}[1]{Fig.~\ref{#1}}

\def\simlt{\mathrel{\lower .3ex \rlap{$\sim$}\raise .5ex \hbox{$<$}}}
\def\simgt{\mathrel{\lower .3ex \rlap{$\sim$}\raise .5ex \hbox{$>$}}}

\begin{document}
\title{Entanglement of condensed magnons via momentum-space fragmentation}
\date{\today}
\author{Clement H. Wong and Ari Mizel}
\begin{abstract}
A scheme is presented for engineering momentum-space entanglement of fragmented magnon condensates. We consider easy plane frustrated antiferromagnets in which the magnon dispersion has degenerate minima that represent ``umbrella" chiral spin textures.  With an applied magnetic field, we tune the Hamiltonian near a quantum critical point that is is signaled by a singularity in the entanglement entropy.  The ground state develops momentum-space entanglement of the chiral spin textures.  The size of the entangled superposition is accessible experimentally through the magnetic structure factor.  Our model is motivated by equilibrium magnon condensates in frustrated antiferromagnets such as CsCuCl$_3$, and it can also be simulated in spin-orbit coupled Mott insulators in atomic optical lattices and circuit quantum electrodynamics.   
\end{abstract}
\maketitle

\textit{Introduction}--Macroscopic quantum coherence has been studied extensively in magnetic and superconducting materials since the early 90's \cite{friedmanNAT00,*chudnovsky05,*wernsdorferPRL97,*braunPRB97,*gargPRL89}. Besides inspiring fundamental interest, macroscopically entangled states have applications in quantum information and metrology.  A well-known example is the N00N state that enables Heisenberg-limited interferometry, motion and magnetic field sensing \cite{chenPRL10,*jonesSC09}, and quantum error correction against photon loss \cite{bergmannPRA16}.
Bose Einstein condensates (BEC's) are natural systems in which to study macroscopic entanglement due to the large number of particles in the ground state.  Entanglement of spatially separated BEC's in optical lattices of ultracold atoms has been achieved \cite{esteveNAT08,*leungMP12,*lauPRL14,*ciracPRA98}.  It was theoretically proposed that the ground state of spin-orbit coupled condensates could support a momentum-space N00N state \cite{stanescu08}. Coherence of counterpropagating quantized superfluid flows in optical lattice rings was also proposed \cite{nunnenkampPRA08,*hallwoodPRA10,*hallwoodPRA11}.  In the solid state, coherent superposition of supercurrent states is the basis for superconducting flux qubits \cite{friedmanNAT00,*korsbakkenPS09,*korsbakkenPRA07,*marquardtPRA08}. In this paper, we study magnon BECs and argue that they can exhibit particularly striking entanglement phenomena.

Researchers are actively investigating quantum effects of magnons in magnetic insulators \cite{zhangCM15,*nakataPRB14,*tabuchiSCI15,*tabuchiPRL14}.  This effort is motivated in part by potential quantum information applications from microwave to optical transducers to robust quantum memories.   Furthermore, due to their bosonic nature, magnons can undergo condensation \cite{nowik-boltyk12,*demokritovNAT06,*tupitsynPRL08,zapfRMP14}. 
%A recent experiment \cite{nowik-boltyk12} measured spatial coherence in the magnon condensate wavefunction corresponding to a standing longitudinal spin density wave \footnote{That work did not show entanglement between condensates; their standing wave interference is consistent with a \emph{product} coherent state  $\ket{PD}=\ket{\beta_{-K}\beta_K}$, where $\ket{\beta_k}=e^{|\beta_k|^2/2+\beta_k b_k^\dag}\ket{0}$, where  $\beta_k=\sqrt{n_{ck}}e^{i\varphi_k}$ and $k=K,-K$ \cite{liSR13}.  In contrast, an entangled condensate state is exemplified by the state $(\ket{\beta_{-K},0}+\ket{0,\beta_K})/\sqrt 2$.}.  

\begin{figure}[t]
\includegraphics[width=\linewidth]{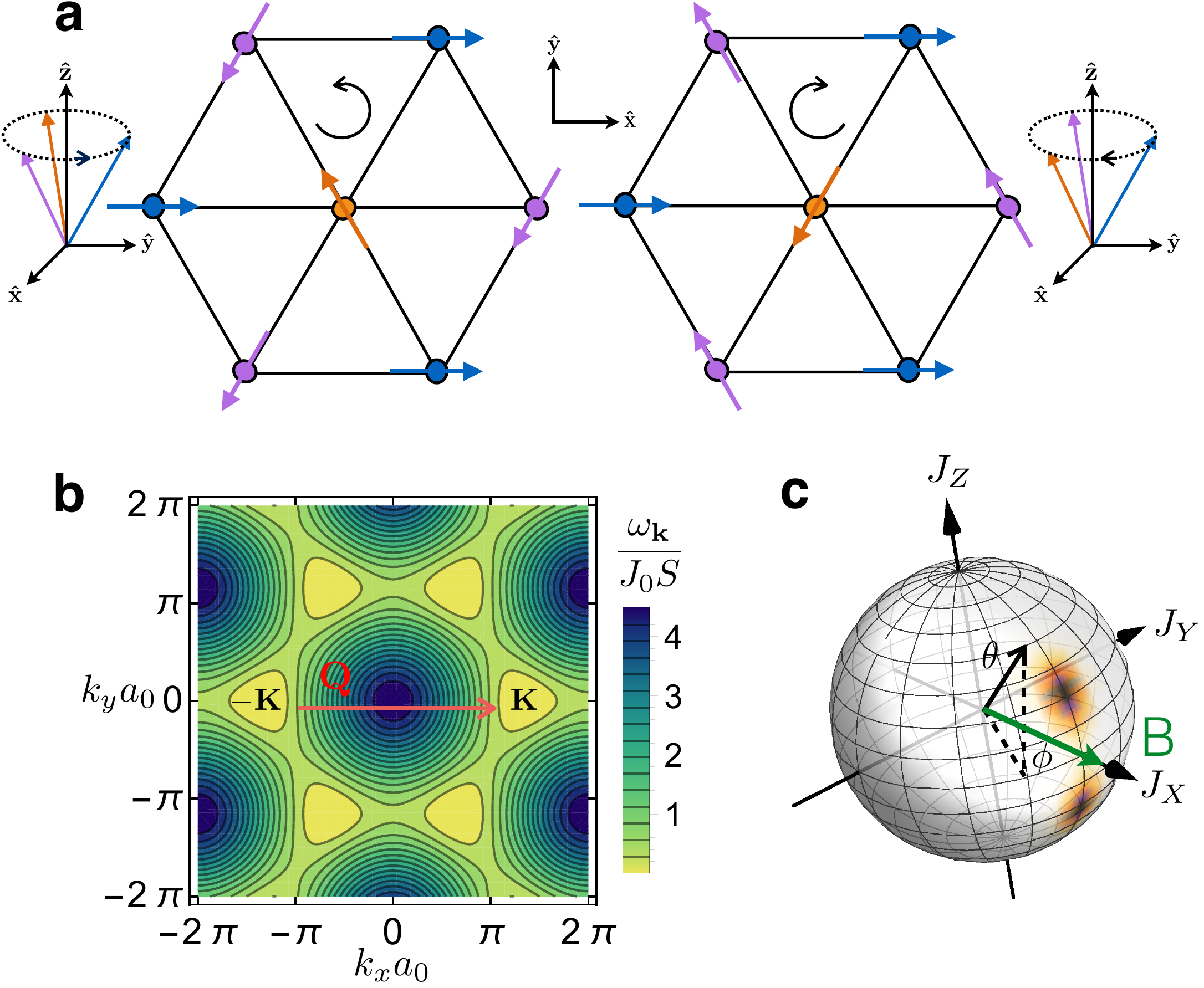}
\caption{(a) Chiral ``umbrella" spin textures in the degenerate ground states of antiferromagnets on the triangular lattice at zero field.
(b) Magnon dispersion with distinct degenerate minima at $\pm \mbK$ {carrying} opposite chiralities.
(c) Superposition of the degenerate ground states represented  on the Bloch sphere (density plot).  North and south poles correspond to magnons completely condensed at $-\mbK$ and $\mbK$, respectively.  Shown here is the state  engineered by coupling the valleys with a sinusoidal field (green arrow)  along $\vu z$ with wave vector $\mbQ=2\mbK$.}
\label{fig:1}
\end{figure}

In this paper, we investigate magnon condensates fragmented into two modes at degenerate minima (valleys) in the magnon dispersion of magnetic insulators \cite{liSR13,leggettRMP01}.  This type of fragmented equilibrium BEC can arise in many frustrated antiferromagnetic insulators with degenerate ground states such as CsCuCl$_3$,  Cs$_2$CuCl$_4$, and Ba$_3$Mn$_2$O$_8$, and Ba$_3$CoSb$_2$O$_9$ \cite{nikuniJPSJ95,*nikuniJPSJ93,zapfRMP14,raduPRL05,*yamamotoPRL14,*coldeaPRL02,*kamiya14}.  Specifically, we consider a  quasi-two-dimensional  canted-XY antiferromagnet (AFM) on a triangular lattice, where easy-plane anisotropy favors  chiral ``umbrella" type ground state spin textures that can be represented as magnon BEC's.  This lattice naturally appears in the planar spin structure of several quantum magnets and can be simulated with engineered spin orbit-coupled Mott insulators in atomic optical lattices \cite{altmanNJP03,*hePRA12,*simonNAT11,*wongPRA13,*wongPRL13} or circuit quantum electrodynamics \cite{houckNAT07,*hurCRP16,*schiroPRL12}.   We propose a method to produce an entangled superposition of magnon BECs in this system and argue that the superposition should be remarkably robust. 

\textit{Magnon condensates}--{The XXZ spin $S$ Heisenberg Hamiltonian on the triangular lattice is given by \cite{nikuniJPSJ95,*nikuniJPSJ93,zapfRMP14} 
\[H_0=J_0\sum_{\mbr,\nu}\qty[S_{\mbr}^xS_{{\mbr+\bm\de_\nu}}^x+S_{\mbr}^yS_{{\mbr+\bm\de_\nu}}^y+\eta S_{\mbr}^zS_{{\mbr+\bm\de_\nu}}^z]+B_0\sum_{\mbr}S_{\mbr}^z\]
where $\mbS_\mbr$ is the spin  operator at $\mbr$,  $J_0$ and $\eta J_0$ are the transverse and longitudinal antiferromagnetic exchange interactions, $\mbr$ is a Bravais lattice vector,  and ${\bm\de_\nu=a(\cos\theta_\nu,\sin\theta_\nu)}$  are unit vectors along nearest neighbor bonds with $a$  the lattice constant and $\theta_\nu=\nu\pi/3$.  We have defined
$B_0=g\mu_BH^e_0$, where $H^e_0$ is the applied external magnetic field, $g$ is the g factor, and $\mu_B$ the Bohr magneton.

Restricting our attention to the case $S>1/2$, we map to a system of bosons using the  Holstein-Primakov  transformation $S_\mbr^-=S_\mbr^x-iS_\mbr^y=\sqrt{2S-n_\mbr}b_\mbr$ and $S^z_{\mbr}=n_\mbr-S$, where $b_\mbr$ are the magnon field operators that satisfy $[b_\mbr,b^\dag_{\mbr'}]=\de_{\mbr\mbr'}$, and $n_\mbr=b^\dag_\mbr b_\mbr$ is the magnon number operator.  In momentum space, $H_0$ to quartic order in magnon operators, becomes ($\hbar=1$)
\[H_0=\sum_\mbk(\w_\mbk-\mu)n_\mbk+{1\over2N}\sum_{\mbk\mbk',\mbq}v_\mbq(\mbk,\mbk') b^{\dag}_{\mbk-\mbq} b_{\mbk'+\mbq}^\dag b_\mbk b_{\mbk'}\]
where $N$ is the number of lattice sites, $b_\mbk$ are destruction operators, $\mu=B_s-B_0$ is the effective chemical potential,  and $B_s=6J_0S(1+2\eta)$ is the saturation field. 
The  $b_\mbk$'s are defined by   $b_\mbr=\sum_\mbk b_\mbk e^{i\mbk\cdot\mbr}/\sqrt{N}$ and satisfy $[b_\mbk,b_{\mbk'}^\dag]=\de_{\mbk\mbk'}$ with $n_\mbk=b^\dag_\mbk b_\mbk$.  The Fourier component of the two-body repulsive density-density interaction has the form $v_\mbq(\mbk,\mbk')=J_0(2\eta\g_\mbq-\g_\mbk-\g_{\mbk-\mbq})$ \cite{kittelQTS87}.  The dispersion is given by $\w_\mbk=2J_0S(3+\g_\mbk)$, where $\g_\mbk=\sum_\nu e^{-i\mbk\cdot\bm{\de}_\nu}$. 
As shown in \fig{fig:1}(b), the dispersion has  two inequivalent degenerate minima (valleys).  We choose to index them by $\pm \mbK=\pm(4\pi/3a,0,0)$.  When $B_0>|B_s|$  or $\mu<0$,  the system's ground state is in the normal phase with fully polarized spins.  We will be interested in the case $B_0\simlt|B_s|$ or $\mu\simgt0$, when the spins are canted out-of-plane, and one can treat the ground state as a Bose condensate of a dilute Bose gas \cite{nikuniJPSJ93,zapfRMP14}.

An approximation to the total Hamiltonian $H_0$ is obtained by projecting onto the valley states to obtain \cite{liSR13} 
\begin{align}
H&=-(\mu+{\chi_1\over2})(\hat n_{-\mbK}+\hat n_\mbK)\nn
&+{\chi_1\over2}(\hat n_{-\mbK}+\hat n_\mbK)^2+(\chi_2-\chi_1)\hat n_{-\mbK}\hat n_\mbK
\label{H}
\end{align}
where $\hat n_i=b^\dag_ib_i$, ${\chi_1=v_{\mathbf{0}}(\mbK,\mbK)/N}$, and $\chi_2=v_\mbQ(-\mbK,\mbK)/N+v_{\mathbf{0}}(\mbK,\mbK)/N$ are the self-interaction and mutual interaction strengths, respectively.  We find that $v_\mbQ(-\mbK,\mbK)=6J_0(1-\eta)$ and $v_{\mathbf{0}}(\mbK,\mbK)=6J_0(1+2\eta)$.  The self-interaction tends to condense magnons into both valleys equally,  leading to a ``fragmented" BEC in momentum space \cite{leggettRMP01}.  The mutual interaction tends to condense magnons into one valley or the other.  We consider the case of easy plane anisotropy, $\eta <1$, so that $\chi_2>\chi_1$, and the mutual interaction is stronger \footnote{Quantum corrections to the interaction strengths are suppressed by $1/S$ \cite{nikuniJPSJ93,nikuniJPSJ95}}.

The leading order ground state energy is obtained by replacing $b_\mbk$ with a condensate wave function $ \ev{b_\mbk}$.  The ground state is doubly degenerate, consisting of a BEC occupying either $\mbK$ or $-\mbK$.   The two ground states exhibit ``umbrella"-type  spin textures $\ev{\mbS}=\sqrt{S(1-B_0/B_s)}[\cos(\mbK\cdot\mbr)\vu x\pm\sin(\mbK\cdot\mbr)\vu y]- S (B_0/B_s)\vu z$, where neighboring spins on triangular plaquette have relative in-plane angles of 120$^o$ as shown in \fig{fig:1}(a).  Focusing on a given triangular plaquette, one sees that the two ground states exhibit spin textures of opposite chirality.

\textit{Engineering entanglement of magnon condensates}-- In this paper, we propose engineering quantum superpositions of these two opposite chirality states by introducing a coupling between valleys.  Physically, this is achieved via  a sinusoidal external magnetic field $\vb H^e_{\mbr}=H^e\cos(\mbQ\cdot\mbr)\vu z$ with wave vector $\mbQ=2\mbK$.  
Up to a  constant, the Zeeman energy adds  $H_B=-B(b^\dag_{-\mbK} b_\mbK+b^\dag_\mbK b_{-\mbK})/2$ to \eq{H},   where $B = g \mu_B H^e$.

Since the total condensate particle number $n_{-\mbK}+n_\mbK\equiv2J$ is a constant of motion, we analyze $H+H_B$ with fixed $J$.
To describe quantum coherence between valleys, it is useful to formally regard $\ket{-\mbK}$ and $\ket{\mbK}$ as  pseudo-spin up  and down, respectively.   We introduce the total valley pseudo-spin operator using the Schwinger representation for angular momentum $\mbJ=\vec b^\dag{\bm\s}\vec b/2$, where $\bm\s$ is the vector of Pauli matrices and $\vec b=(b_{-\mbK},b_\mbK)$.  
The operators $\mbJ$ satisfy the usual angular momentum algebra with $\mbJ^2=J(J+1)$ \cite{sakuraiQM}.  
The valley polarization operator $\hat J_Z=(b_{-\mbK}^\dag b_{-\mbK}-b_\mbK^\dag b_\mbK)/2$ defines the eigenstates  $J_Z\ket{J, m}=m\ket{J, m}$ where $m=(n_{-\mbK}-n_\mbK)/2$ and $-J\leq m\leq J$.  We define the pseudo-spin  Hamiltonian by $H+H_B=H_J-2J(\mu+\chi_1/2)-J^2(\chi_1+\chi_2)$, where
\ben
H_J=-AJ_Z^2-B J_X~,\label{Hj}
\een
and $A=\chi_2-\chi_1=v_\mbQ(-\mbK,\mbK)/N$.

The Hamiltonian \eq{Hj}, viewed as a many spin system with infinite coordination number, is a limit of the Lipkin-Meshkov-Glick \cite{dusuelPRB05,latorrePRA05,barthelPRL06,botetPRB83} model.\footnote{The terms  $J_X^2,J_Y^2$ are absent since they are prohibited here by momentum conservation.}  In the thermodynamic (TD) limit $J\to\infty$, \eq{Hj} exhibits a second order quantum phase transition (QPT) at the critical field $B_c=2AJ$ between eigenstates of $-J_Z^2$ and $-J_X$, which are distinguished by the order parameter $\ev{J_X}$.
\footnote{This transition bears similarity to the QPT between the normal and superradiant phase in the Dicke model \cite{yixiaoCTP15}.}
In the  ``broken" phase, at $B<B_c$,  $\ev{J_X}=B/B_c$, and the ground state is doubly degenerate and gapless.
In the  ``symmetric" phase, at $B\geq B_c$, $\ev{J_X}=J$, and the ground state is nondegenerate and gapped.

\begin{figure}[t]
\includegraphics[width=\linewidth]{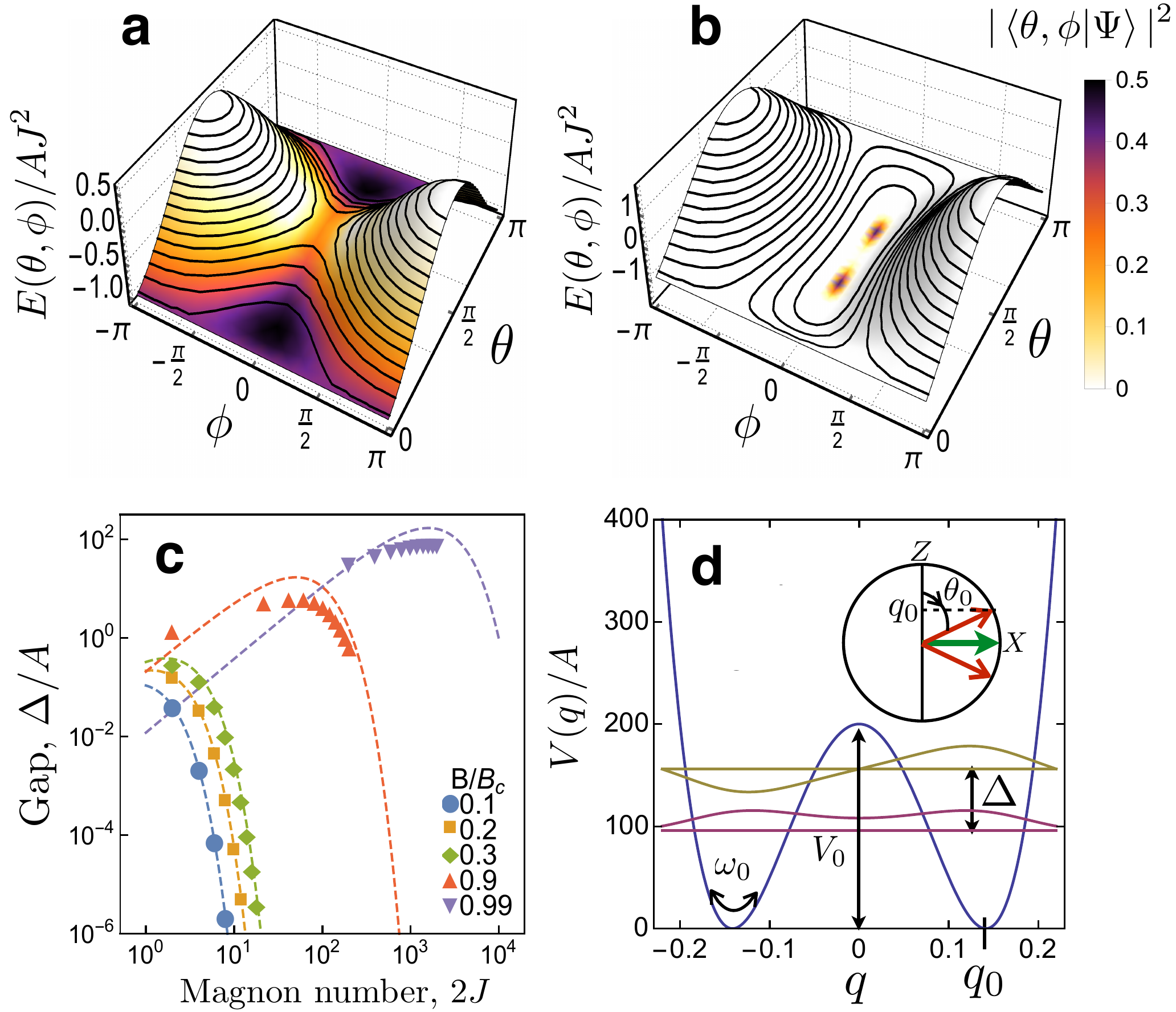}
\caption{Top panel: (surface plot) Mean field ground state energy $E(\theta,\phi)$ given in \eq{Ej}. (color density plot) Ground state probability density  expressed in the pseudo-spin coherent state basis given in \eq{w}. In (a) $(B/B_c,J)=(0.3,3)$ and (b) $(B/B_c,J)=(0.9,10^2)$. 
Bottom panel: 
(c) Tunnel splitting in units of the interaction strength, $\D/A$, as a function of magnon number $2J$. Solid lines are computed by numerical diagonalization of \eq{Hj} and dashed lines are computed from instanton formulae.
(d) Effective quartic double well potential, {$V(q)/A$}, in the polar pseudo-spin angle $q=\cos\theta$ in the regime $\e\ll1$ with $(\e,j)=(10^{-2},10^{3})$. Ground and excited state wavefunction computed from numerical diagonalization are also plotted.}
\label{fig:2}
\end{figure}

Rather than focusing on the thermodynamic limit, we  seek a finite size system with (i) an energy gap to excitation sufficiently large to observe the ground state and (ii) a ground state that manifests macroscopic entanglement.  Working in the $J_Z$ basis $\ket{J, m}$, we first numerically compute the ground state $\ket{\Psi}=\sum_Nc_{Jm}\ket{J, m}$ and energy gap of \eq{Hj}.  To develop an understanding of the results, we go to the spin coherent state basis
\ben\ket{\theta,\phi}={1\over(1+|w|^2)^{J}}\sum_{m=-J}^{J}{{2J}\choose{J-m}}^{1/2}{w^{J-m}}\ket{J,m}\label{w}\een
where $w=e^{i\phi}\tan{\theta/2}$ is the stereographic projection of the sphere onto the complex plane. Since \eq{w} is an eigenstate of $\mbn\cdot\mbJ$ with $\mbn=(\sin\theta\cos\phi,\sin\theta\sin\phi,\cos\theta)$, it can be visualized as a point on the pseudo-spin Bloch sphere, as shown in \fig{fig:1}(c). The ground state can be completely characterized \cite{sandersPRA89} by the probability density $\ev{\rho}{\theta,\phi}$,  where $\rho=\ketbra{\Psi}$ is the density matrix.  Results are shown in \fig{fig:1}(b) and \fig{fig:2}(a)-(b).

Numerical results for the gap are plotted in \fig{fig:2}(c).  The gap arises from tunneling between degenerate mean field ground states.  They can be approximated by taking \eq{w} as a variational ground state of $H_J$ with $(\theta,\phi)$ as parameters.  The value of $\ev{H_J}{\theta,\phi}$ is 
\ben E(\theta,\phi)\equiv-AJ^2\qty(\cos^2\theta+2{B\over B_c}\sin\theta\cos\phi),\label{Ej}\een 
neglecting smaller terms of order $J$.  Degenerate minima of \eq{Ej} occur at $(\theta,\phi)=(\theta_0,0)$ and $(\theta,\phi)=(\pi-\theta_0,0)$ where $\sin\theta_0 = B/B_c$.  The minima are separated by a tunnel barrier in the $\theta$ direction with height $V_0=E(\pi/2,0)-E(\theta_0,0)$. The transition amplitude $\mel{\theta}{J_X}{\pi-\theta}$ causes tunneling between the degenerate minima, leading to a ground state that is a symmetric superposition separated in energy from the antisymmetric superposition by a tunneling splitting.  The value of the tunneling splitting can be computed using instanton methods \cite{gargPRB92} based on the path integral representation of the propagator in the \eq{w}  basis
$\mel{\theta_f,\phi_f}{e^{-i H_J t}}{\theta_i,\phi_i}=\int_{(\theta_i,\phi_i)}^{(\theta_f,\phi_f)}\mcal D\theta \mcal D\phi\exp\qty[i\int_0^t dt' \qty(J\dot{\phi}(1-\cos\theta)-E(\theta,\phi))]$.
To elucidate the results, we set $\e = 1-B/B_c$ and separately consider $\e \approx 1$ and $\e \ll 1$.

Far below the QPT, when $B \ll B_c$ and $\e \approx 1$, one finds that $\sin\theta_0\ll1$, so that $V_0$ presents a high tunnel barrier in $\theta$.  This leads to N00N-like  ground states, well-localized near the poles  as shown in \fig{fig:2}(a), but with a very small tunnel splitting  \cite{chudnovskyPRL88,garanin91,chudnovsky05} $\D=A(4 J^{3/2}/\sqrt{\pi})(e/2)^{2 J}(B/B_c)^{2 J}$. The $2J$ power law dependence on  $B/B_c$ can be understood from perturbation theory, since the degeneracy of the $J^2_Z$ eigenstates is lifted by the perturbation $BJ_X$ in the  $2J^{\rm th}$ order. As shown \fig{fig:2}(c), the tunnel splitting ${\D/ A}<1$ is exponentially suppressed with magnon number.  While this N00N-like state exhibits the entanglement that we seek, it will be difficult to prepare and observe since the tunnel splitting is too small.

To attain a larger tunnel splitting, we take $B$ close to $B_c$, so  that $\e\ll1$.  We find that $\theta_0\approx\pi/2-\sqrt{2\e}$.  The classical minima at $(\theta,\phi)=(\theta_0,0)$ and $(\theta,\phi)=(\pi-\theta_0,0)$ therefore approach one another and the tunneling barrier height $V_0= A(J\e)^2$ decreases.  Quantum fluctuations in $\phi$ are strongly suppressed, as shown in \fig{fig:2}(b), and  one can integrate out $\phi$.  The result is an effective Lagrangian $L={m\dot q^2/2}-V(q)$, where $q\equiv\cos\theta$,  $m=1/2A$ is an effective mass, and $V(q)=V_0(q_0^2-q^2)^2/q_0^4$ is a quartic double well potential with minima at $\pm q_0=\pm \sqrt{2\e}$ \cite{gargPRL89,gargPRB92}.  This potential appears in {\fig{fig:2}(d)}, together with its ground and excited state wavefunctions. The tunnel splitting is $\D=4\sqrt{3}{\w_0}\sqrt{S_0/2\pi}e^{-S_0}$, where $S_0=(2J/3)(2\e)^{3/2}$ is the instanton action and $\w_0=2JA\sqrt{2\e}$ is the attempt frequency \cite{gargPRL89}.   
This tunnel splitting is compared to the numerical diagonalization in \fig{fig:2}(c).   For fixed $\e$, we can scale $J$ to maximize the gap: $J = J_{\rm max}(\e)=0.8\epsilon^{-3/2}$.  This yields $\D_{\rm max}=12J_0(1-\eta){J_{\rm max}^{2/3}/N}$.  Choosing $\e=10^{-2}$, $N=10^4$ lattice sites, and  $2J = 10^3$ magnons, and noting typical values $\eta=0.8$ and $J_0=5$ K \cite{nikuniJPSJ95}, we find a splitting of $\D=120$ mK.  This should be sufficiently large to permit initialization of the ground state in dilution refrigerator temperatures of $15$ mK.

\textit{Measures of momentum-space entanglement}-- With the scaling $J = J_{\rm max}(\e)=0.8\epsilon^{-3/2}$, the states localized at the minima $\pm q_0$ remain distinct with $\ev{J_Z}\sim \pm J q_0 = \pm J_{\rm max}(\e)q_0\sim \e^{-1}$,  even though the distance between minima goes to zero as $2q_0 = 2 \sqrt{2 \e}$.   Thus, the entanglement does not vanish as $\e$ shrinks.
We can measure this entanglement using the entanglement entropy, given by the von Neumann entropy of the reduced density matrix of $2j$ magnons $S_E(J,j)=-\tr[\rho_{2j}\log_2\rho_{2j}]$, where
\[(\rho_{2j})^{k_1\ldots k_{2j}}_{l_1\ldots l_{2j}}\equiv (2(J-j)!/2J!)\ev{b^\dag_{k_1}\ldots b^\dag_{k_{2j}}b_{l_1}\ldots b_{l_{2j}}}\]
with $k_i, l_i$ indices in any basis  \cite{korsbakkenPRA07}. The entanglement entropy for several values of $J$ and $j$ is plotted in  \fig{fig:entangle}(a).  At $\e=1$, $S_E=1$ as expected for a N00N state, where mainly two states are occupied, while  $S_E\to0$ at $\e\ll0$, where the ground state approaches an eigenstate of $J_X$. Near the quantum critical point $B\to B_c$, where we propose working, $S_E$ remains near 1.  In fact, a cusplike peak is apparent \cite{dusuelPRB05,latorrePRA05} that has the form of a logarithmic divergence in the thermodynamic limit \cite{barthelPRL06}.

The behavior of $S_E$ confirms the presence of entanglement in our ground state, but it is an imperfect measure.  Its value is affected by entanglement due to symmetrization of the wave function -- $S_E$ is non-zero even for a single state $\ket{J, m}$.
The ground state of our system is a superposition of states with large differences in $\ev{J_Z}$, so a suitable measure \cite{marquardtPRA08} of the amount of entanglement is the variance $\langle\D\hat J_Z^2\rangle$, where $\D \hat J_Z \equiv \hat J_Z-\langle{\hat J_Z}\rangle$.  For the parameter values listed above, $\e=10^{-2}$ and $2J=1000$, we have $\langle\D\hat J_Z^2\rangle=58.9$.  To study the behavior of the variance as a function of $B/B_c$, we define \cite{morimaePRA05} the scaling exponent $p$ by $\sqrt{\langle\D\hat J_Z^2\rangle} =\order{J^p}$.    \fig{fig:entangle}(a) plots a numerical fit to $p$.  We find that $p=1$ in the range $0<B/B_c\simlt0.9$, and $p=0.5$ at $B/B_c\gg1$, where the ground state $\ket{\theta=\pi/2,\phi=0}$ is separable when written in terms of $b^{\dagger}_\mbK$ and $b^{\dagger}_{-\mbK}$. For small $\e = 1-B/B_c$, $p$ goes as ${1.15+0.16\log \epsilon}$ as shown in the inset of \fig{fig:entangle}(b).  Thus, $B/B_c\leq0.99$ implies $p>0.7$, which we regard as the entanglement region. 

It is appropriate to define the entanglement region in terms of $p$ because $p$ determines the ability of the system to find use in precision quantum metrology.  If a relative phase is accumulated between $\ket{\pm\mbK}$, leading to the state $\ket{\Psi(\phi)}=e^{i\phi J_Z}\ket{\Psi}$,  measurement of this phase %  by applying a detuning $\w_ZJ_z$from circulating B field?
 will have a minimum phase estimation error $\de\phi$ bounded by the quantum Cramer-Rao bound $\de\phi_{\rm min}=1/\sqrt{F_Q}$, where $F_Q$ is the quantum Fisher information \cite{braunsteinPRL94}. % From theory of quantum parameter estimation 
For pure states  $F_Q=\ev{\D J_Z^2}$, so that $\de\phi_{\rm min}\propto J^{-p}$.  Therefore, $p$ measures the scaling of precision with magnon number; $p=0.5$ is the so-called standard quantum limit, while $p=1$  is the Heisenberg limit, showing quantum enhanced precision. 

For experimental measurement, the $J_Z$ variance can be related to a spin correlation function that can be probed with neutron scattering \cite{kittelQTS87}
\[\ev{\D J_Z^2}=J(J+1)-\ev{J_X^2}-\ev{J_Y^2}={J^2}-N\ev{S^z_\mbQ S^z_{-\mbQ}}\]
where we used $\ev{J_Z}=0$.  The order parameter $\ev{J_X}$ of the QPT appears in the density of the longitudinal spin density wave%(condensate magnon density)
\[\ev{\hat n_\mbr}={2J}+2{\ev{J_X}}\cos\mbQ\cdot\mbr\] 
which can be measured by Brillouin light scattering \cite{nowik-boltyk12}. 

\begin{figure}[t]
\includegraphics[width=\linewidth]{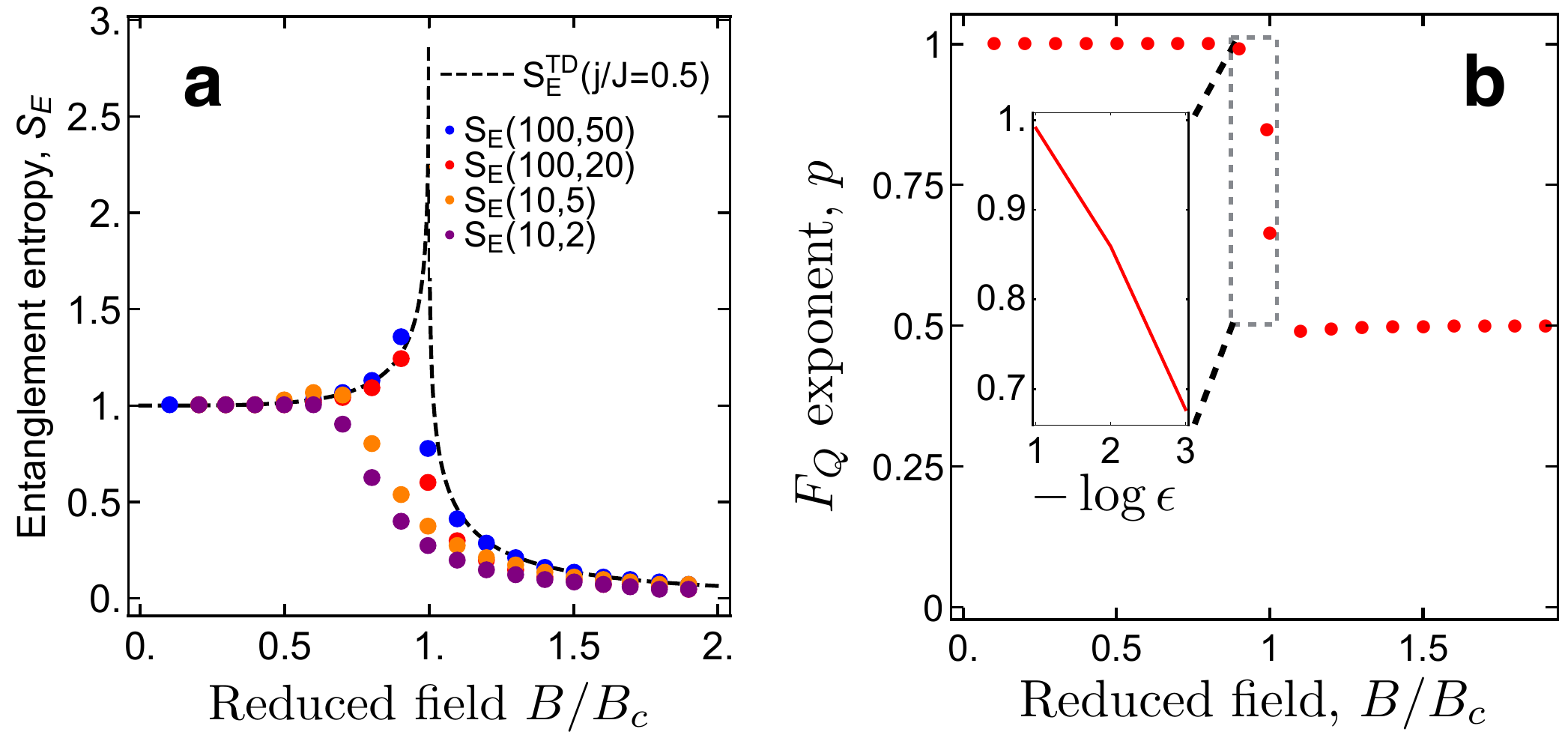}
\caption{(a) Momentum space entanglement entropy $S_E$ as a function of the reduced field $B/B_c$.  Dots show numerically computed values. Dashed line shows theoretical curve in the thermodynamic limit.  (b) The exponent $p$ of the quantum Fisher information $F_Q$ as a function of $B/B_c$.  Inset shows logarithmic plot  in the crossover region $\e=1-B/B_c\leq 0.1$.}
\label{fig:entangle}
\end{figure}

\textit{Decoherence}-- The Hamiltonian \eq{Hj} is formally identical to a model for a uniaxial ferromagnet in a transverse field  \cite{chudnovskyPRL88,gargPRL89,garanin91}. However, in that case, coherence is highly vulnerable to dephasing from Zeeman coupling to low frequency magnetic field noise \cite{chudnovsky05}.  For a system starting in the ground state, such noise will lead to excitations via a large $J_Z$  matrix element to the first excited state.  In contrast, one of the remarkable aspects of this magnon system is its robust coherence.  Ambient magnetic field noise perturbs the pseudo-spin $J_X$ term in the magnon Hamiltonian, which should have a negligible effect on the ground state because its $J_X$ matrix element to the first excited state is very small.
A $J_Z$ dephasing term would require a highly non-trivial interaction capable of differentiating  between  spin  textures  of  opposite chirality.  
   
Instead, the lifetime of the magnon quantum state is determined by magnon loss.  Uniaxial $U(1)$ symmetry breaking terms arise from magnetic anisotropy, Dzyaloshinsky-Moriya interactions,  and dipolar interactions. %of the form $S_{x}S_z$ and $S_{y}S_z$,
The magnitude of the dipolar interaction, which is the dominant term, scales as $V_d={g^2\mu_B^2/ a_0^3}$ \cite{tupitsynPRL08}. For a typical lattice constant $a_0=0.7$ nm \cite{adachiJPSJ80}, this gives $V_d\sim 10$ mK. 
As a result, the condensate can lose magnons by spontaneous emission into the magnon bath at the rate $\al=|v_d^{(3)}|^2D$, where $v_d^{(3)}\sim V_d/\sqrt{N}$ is the three-magnon scattering amplitude, 
$D=\mcal Am/\pi $ is the bath density of states, $m\sim 1/J_0 a_0^2$ is the  magnon effective mass, and $\mcal A\sim Na_0^2$ is the area.  The magnon loss rate  can thus be estimated as $\al\sim V_d^2/J_0 \pi \sim 10^{-2}$mK, which is much slower than time scale set by the gap $\D$.  Moreover, the form of the ground state of our system should be much more robust against particle loss than, say, a N00N state \cite{dornerPRL09,*huangSR15}. 

\textit{Outlook}--This work presents a proposal for establishing momentum-space entanglement of condensed magnons in a quasi-two-dimensional canted XY antiferromagnet  on a triangular lattice. The  entanglement explored here may also be present in a quasi-equilibrium magnon BEC in yttrium-iron garnet (YIG), which warrants further study.
% \cw{suggest implementation of ``0-$\pi$" qubit}
%The power law dependence of this gap on the external magnetic field at wave vector $2K$ can be used for magnetic field sensing.
%The magnons studied in this work couple to circularly polarized light, which potentially enables transfer of the entangled magnon state to photons or photon-mediated coupling to a superconducting qubit \cite{tabuchiSCI15}.
%The one-dimensional interacting bosonic system studied here can be solved using Luttinger liquid techniques, and in particular, analytic formulae for the ground to excited state gap including effects of noncondensate states can be obtained, which we relegate to future work. 

% and has the advantage controllability via spSHE
%\cite{kajiwaraNAT10},
 %In this work, we focused on the condensate states and  has been to neglect noncondensate 
% persistent magnon currents 1D ring, control via magnetic texture induced Berry phase or AC effect
 
\bibliography{physics}

\end{document}